\documentclass[twocolumn]{aastex63}

\usepackage{natbib}

\received{datum, 2021}
\revised{datum, 2021}
\accepted{datum, 2021}

\submitjournal{ApJ}

\shorttitle{Statistical Study of Propagating Kink Waves}
\shortauthors{Tiwari et al.}

\begin{document}

\title{A statistical study of propagating MHD kink waves in the quiescent corona}

\correspondingauthor{Ajay K. Tiwari}
\email{ajaynld13@gmail.com}
\author[0000-0001-6021-8712]{Ajay K. Tiwari}
\affil{Northumbria University \\
Newcastle upon Tyne, NE1 8ST, UK}
\affiliation{Centrum Wiskunde \& Informatica, Amsterdam, 1098 XG Amsterdam, The Netherlands}

\author[0000-0001-5678-9002]{Richard J. Morton}
\affiliation{Northumbria University \\
Newcastle upon Tyne, NE1 8ST, UK}
\author[0000-0002-7863-624X]{James A. McLaughlin}
\affiliation{Northumbria University \\
Newcastle upon Tyne, NE1 8ST, UK}
\newcommand{\comp}{CoMP}
\begin{abstract}
The Coronal Multi-channel Polarimeter (CoMP) has opened up exciting opportunities to probe transverse MHD waves in the Sun's corona. The archive of CoMP data is utilised to generate a catalogue of quiescent coronal loops that can be used for studying propagating kink waves. The catalogue contains 120 loops observed between 2012-2014. This catalogue is further used to undertake a statistical study of propagating kink waves in the quiet regions of the solar corona, investigating phase speeds, loop lengths, footpoint power ratio and equilibrium parameter values. The statistical study enables us to establish the presence of a relationship between the rate of damping and the length of the coronal loop, with longer coronal loops displaying weaker wave damping. We suggest the reason for this behaviour is related to a decreasing average density contrast between the loop and ambient plasma as loop length increases. The catalogue presented here will provide the community with the foundation for the further study of propagating kink waves in the quiet solar corona.

\end{abstract}


\section{Introduction}\label{intro}
The presence of MHD waves in the solar atmosphere is now well established \citep[e.g.,][]{1997ApJ...491L.111O, 1998A&A...337..287E, forest_gurman,  asch1999, naka1999,1999SoPh..187..261S, 2000A&A...355L..23D, 2001MNRAS.326..428W, 2002MNRAS.336..747W,goossens2002, 2002A&A...393..649M, 8007,2003A&A...406..709K, 2003A&A...406.1105W, 2005A&A...430L..65V, tom2007, tom2009, rick2012, MORetal2016, MORetal2019}. Of all the MHD wave modes observed in the solar atmosphere, some of the most interesting are the transverse waves. They are thought to be critical in transferring energy from the turbulent convection in the photosphere to the solar corona.

The most common transverse wave in the corona appears to be the kink mode (the presence of torsional modes are more difficult to determine; however, there is some evidence for such motions, e.g. \citealt{verwichte2020}). The kink mode has, to date, {{been observed in the corona}} in three variants: {{\textit{(decaying) standing} kink waves, \textit{\lq{decayless}\rq{} standing} kink waves, and \textit{propagating} kink waves}}.
The standing kink waves were the first transverse wave modes to be observed in active region coronal loops \citep{naka1999, asch1999}, found with the Transition Region and Coronal Explorer \citep[TRACE -][]{TRACE1}. These observations, and the launch of the Solar Dynamics Observatory \cite[SDO -][]{PESetal2012}, heralded a new era in the exploration and understanding of the physical properties of the solar corona through standing kink modes \citep{naka_review_2020}. The standing kink modes are typically observed in active region coronal loops following an eruptive process \citep{stepanov2012, Kinkexcitation15, goddard2016}. The excitation mechanism of these standing kink waves is believed to be nearby eruptions or plasma ejections \citep[rather than a blast shock wave ignited by a flare, as previously thought, e.g.,][]{Kinkexcitation15}, which leads to a displacement of the coronal loops from their equilibrium position. These waves are found to be rapidly damped, with the damping being attributed to the phenomenon of resonant absorption or mode coupling \citep[e.g.][]{ionson1978, holloweg1984, ruderman2002, goossens2002, asch2003}. More recently, \cite{goddard2016} produced a catalogue of standing kink modes observed with the SDO Atmospheric Imaging Assembly \cite[AIA -][]{LEMetal2012}, which was later extended by \cite{Nechaeva_catalogue}. A study of the relationship between damping time and amplitude indicated that a non-linear damping mechanism might also contribute to the observed damping. \cite{Van_Doorsselaere_2021} suggested that the observed relationship could be explained by uni-turbulence, a form of generalised phase-mixing \citep{Magyar_2019, Van_Doorsselaere_2020}.

\smallskip

Secondly, there has been the discovery of `decayless' standing kink wave modes \citep{TIANetal2012, Wang2012,
Anfino2013, Anfino2015, Nistico2013} in active region loops. These low-amplitude ($<1$~Mm) oscillations do not appear to damp in time and are seen for several cycles. In some cases, the wave amplitudes are shown to gradually grow \citep[e.g.,][]{Wang2012}.

\smallskip

Finally, it was demonstrated that there are persistent and ubiquitous fluctuations in the Doppler velocities of coronal emission lines, which propagate at Alfv\'enic speeds and follow magnetic field lines \citep{tom2007, MORetal2019, Yang2020b, Yang2020a}. These motions have been interpreted as propagating kink waves and have also been observed with SDO/AIA \citep[e.g.][]{MCIetal2011, THUetal2014, WEBetal2018, WEBetal2020}. There have been several studies to reveal the properties of the propagating kink waves, finding that the power spectra of the velocity fluctuations can be described with a power law, and also show an enhancement of power at $4$~mHz \citep[e.g.][]{tom2009, MORetal2015, MORetal2016, MORetal2019}. The excitation mechanism for the propagating waves is believed to be the random shuffling of magnetic elements in the photosphere due to convection, although this mechanism appears only to be able to explain the high-frequency part of the observed power spectrum \citep{CRAVAN2005}. Observational and theoretical studies provide evidence which indicates that mode conversion of $p$-modes may play a role in exciting some fraction of the observed waves \citep{cally2017, MORetal2019}. Moreover, the origin of the low-frequency velocity fluctuations is still unclear, although \citet{Cranmer2018} suggest that reconnection resulting from the evolution of the magnetic carpet may be the source. 

The damping and dissipation of the propagating kink waves have not yet received as much attention as the standing modes. To date, there has only been a single observational case study that has been analysed. \cite{verth2010a} were the first to highlight the wave damping of the event presented in \cite{tom2009}, suggesting resonant absorption could provide a reasonable description of the observed behaviour. A number of other studies have also used this event as a case study \citep[e.g.][]{VERetal2013b,PASetal2015,Tiwari_2019,Montes_Solis_2020}.



\smallskip

The focus of many previous studies has been on the {{(decaying and \lq{decayless}\rq{})}} standing kink waves observed in active regions, with many statistical studies revealing the typical properties of these modes \citep[e.g.,][]{Anfino2015, Nechaeva_catalogue}. Given that the quiescent corona occupies a larger volume of the Sun's atmosphere than active regions and is omnipresent over the solar cycle, it is vital to understand the nature of the propagating kink waves that exist there and the waves' role in heating the quiescent coronal plasma. However, to date, there has been little focus on the {{propagating}} kink waves observed in quiescent corona. This paper attempts to fill some of that gap in our knowledge and provides a catalogue of suitable quiescent coronal loops that can be used for studying the propagating kink waves. In generating this catalogue, an overview of some of the typical {propagating} kink wave properties in the quiescent Sun is also provided. This paper also serves as a natural extension to the study by \citet{Tiwari_2019}. 

\medskip 

The paper is structured as follows: In Section~\ref{data} the details of data and the analysis methods used are provided. Section~\ref{result} presents the main results and discusses the findings. The paper is concluded in Section~\ref{conclusion}.


\section{Data and analysis}\label{data}

 \subsection{Observations} \label{comp}
The data are obtained from the Coronal Multi-channel Polarimeter \citep[{\comp} -][]{, tomczyk2008}, a coronagraph which observes the off-limb corona between 1.05~$R_{\Sun}$ and 1.35~$R_{\Sun}$. CoMP is an imaging spectro-polarimeter, and provides images of the corona taken at three different wavelength positions centred on the 10747~{\AA} Fe {\sc{XIII}} coronal emission line (referred to as three-point measurements). The data {were} selected from {\comp} observations taken between 2012-2016. The data sets in which there were more than 135 (near)-contiguous data frames is identified by a manual inspection on the \href{https://mlso.hao.ucar.edu/mlso_data_calendar.php?calinst=comp}{\comp}\footnote{https://mlso.hao.ucar.edu/mlso\_data\_calendar.php?calinst=comp} data web-page. The dates are given in Table~\ref{table01}. The data sets from each selected date have a temporal cadence of 30 seconds (some with a small number of missing frames, $<5\%$) and spatial sampling of $\sim4\farcs$5. The Doppler velocity data products derived from fitting a Gaussian model to the line profile from the three-point measurements are the focus of this study. Details of the procedure used to estimate the Doppler velocities are given in \citet{Tian2013}, and an assessment of their uncertainties is performed in \citet{MORetal2016}. A time-series of Doppler velocity images of the corona is used for this study. In cases where frames are missing, linear interpolation is performed to fill the gaps. Further registration of the Doppler images within each time-sequence is undertaken via cross-correlation, with further details given in \citet{MORetal2016}. 

\medskip

\begin{figure*}[!ht]
    \centering
    \includegraphics[width=1.0\textwidth]{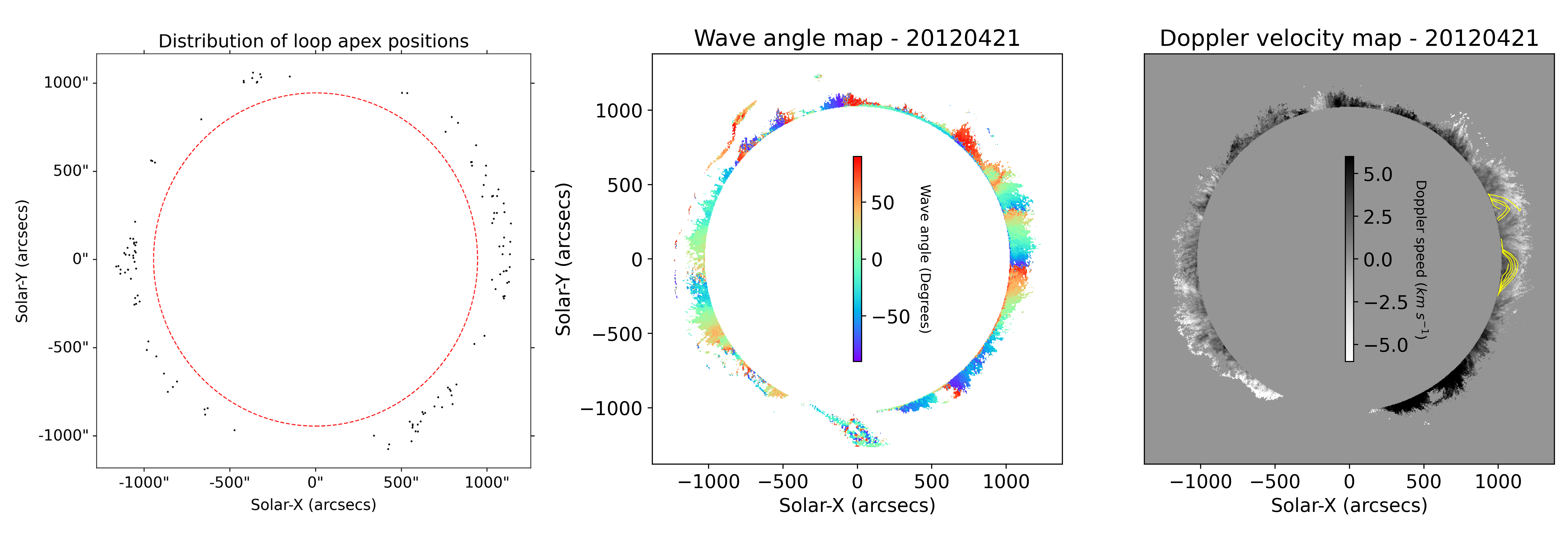}
    \caption{\text{Left panel} shows the position of all loops analysed. The black dots correspond to the position of the loop apex. The red dashed lines represents the solar disk. A sample observation for the loops observed on 19-July-2012. The \textit{Centre panel} shows the calculated wave angle at each pixel position, determined using a coherence based method on each pixel of the Doppler velocity image. This wave angle serves as the guide for the tracks shown in yellow in the \text{Right panel}. }
    \label{figwave-angle-map}
\end{figure*}

The analysis of the {{propagating}} kink waves requires the measurement of the wave propagation direction. Hence, a data product called a \emph{wave angle map} is also derived, which gives the relative direction of propagation for the velocity signal within each pixel. The basis of the wave angle calculation requires a coherence-based approach for the analysis of the velocity signals, with general details discussed in \citet{MCIetal2008} and its use on CoMP data is discussed in \citet{tom2009}, \citet{MORetal2015} and \citet{Tiwari_2019}. The strategy is to use the coherence between the Doppler velocity time-series of each pixel and its neighbouring pixels to obtain islands of coherence above a threshold value. The direction of wave propagation is calculated by a straight line fit through the islands, which minimises the sum of perpendicular distances from the points to the line. Performing this operation for each pixel of the Doppler velocity images gives the wave angle map. A sample wave angle map is shown in the \textit{centre panel} of Figure~\ref{figwave-angle-map}.

\subsection{Selection of loops for study}\label{pfss}
The selection of the loops from the {\comp} data is a critical step in the analysis of the waves. For each of the data sets, suitable systems of coronal loops are identified. A lower limit of 50 Mm is placed on the lengths of the loop systems selected, whereby length refers to their visible length in the CoMP field of view\footnote{An issue with measurements of the absolute length of loops is that the occulter on CoMP covers the solar disk out to $\sim$1.05~$R_\odot$, meaning that it is not possible to trace the coronal loops back to their footpoints. Hence, only the length of the loop in the CoMP field-of-view is reported.}. This limit is required to preserve a high signal-to-noise level in the Doppler velocity time-series. Smaller loops are closer to the occulting disk, where the signal suffers from a high scattering of photons that leads to poor estimates for the Doppler velocity and hence increased noise in time-series. The imposition of a minimum loop length also ensures an appropriate sampling in the $k$-direction in Fourier space, which is required for further analysis. A few such selected loops are shown in the \textit{right panel} of Figure~\ref{figwave-angle-map}. The loops are chosen by manually identifying closed-loop structures first in the Doppler and intensity image sequences and later also in the wave angle map. The closed-loop structures as arcades of loops that appear semi-circular are identified, with each leg starting near the occulting disk. 

The second criteria for loop selection require that the loops should be orientated such that the longitudinal direction of the magnetic field is close to being positioned in the plane-of-sky. The geometry and orientation of the loops are identified by performing magnetic field extrapolations using the Potential Field Source Surface (PFSS - \citealt{derosapfss}) software. PFSS extrapolations performed using line-of-sight magnetogram data obtained from SDO's Helioseismic and Magnetic Imager (HMI - \citealt{Scherrer2012}). The extrapolations provide us with a schematic geometry and orientation of the loops in the plane-of-sky. The extrapolated field lines visibly agree with the loop structures observed in the coronal EUV images obtained by SDO/AIA and with the intensity images obtained by CoMP. 
Furthermore, the loops are selected to avoid loops within the cores of active regions (i.e. rooted in or near sunspots). Some of the trans-equatorial loops identified were located in the extended plage region of active regions on the visible solar disk. The observed loops are assumed to be rooted in network regions and are not part of active region loop systems. The \textit{left panel} of Figure~\ref{figwave-angle-map} displays the location of the apex positions, with respect to the solar limb, of all the loops used within this study.

Due to the low spatial resolution of CoMP, the effect of line-of-sight integration, and projection effects, there is an issue of disambiguation of individual loops in \comp \ observations. Hence, the focus is on the wave signals in the system of coronal loops, as opposed to an individual structure.

\subsection{Wave parameter estimation}\label{wave-angle-calc}

For each system of coronal loops, a number of wave paths are extracted. A \textit{wave path} is defined as being a contiguous set of pixels through the loop system, starting and finishing at the occulting disk. A pixel is selected within the loop system, and the wave angle map is utilised to map out a path, selecting the subsequent pixel based on the angle of propagation. The path is followed until the occulting disk is reached. A square box-car median filter of width two pixels was applied to the wave angle map to try and suppress some of the noise and led to the improved tracing of wave-paths. The pixel locations for each wave path are then used to extract the relevant velocities from the Doppler velocity images for each frame in the image sequence. A cubic interpolation maps the velocities from the selected wave paths onto $(x,t)$ space. For each wave path, the neighbouring five wave-paths on either side of the original wave path are also extracted. The result of this is time-distance diagrams along the coronal loops systems. 
The longer loops lead to additional issues when tracing them because the wave angle suffers from more significant uncertainties closer to the apex of these loops. This arises because the wave angle is being poorly estimated near the upper boundaries of the wave angle map, primarily due to lower signal-to-noise in these regions arising from fainter coronal emission. In such cases, only the wave path for half a loop is obtained.
The half-loop length is defined for each loop, obtained by finding the point of inflexion for the traced trajectory of the wave-path (except for the longer loops where only a half-loop is already traced). 

The Doppler velocity time-distance diagram for each half-loop is subject to a two-dimensional Fourier Transform. The Fourier components are used to separate the inward and outward components of the wave propagation and provide velocity power spectra as a function of wavenumber-frequency ($k-\omega$). 

\medskip
The propagation speed for the waves is calculated in a manner similar to \citet{tom2009}, \citet{MORetal2015}, \citet{MORetal2019} and \citet{Tiwari_2019}. It is straightforward to filter either the inward or outward waves from the Fourier-transformed Doppler velocity time-distance diagrams by setting Fourier components equal to zero. The inverse Fourier transform of the filtered Fourier components then provides a Doppler velocity time-distance diagram containing only the inward or outward propagating waves. The cross-correlation between the time-series at the centre of the wave path and the neighbouring time-series along the path is calculated from these filtered Doppler velocity time-distance diagrams. The location of the peak of the cross-correlation function gives the time-lag between the signals and is determined by fitting a parabola to the peak. The observed lags as a function of the position along the wave path are fit with a linear function, and the gradient gives the propagation speed of the wave. 

\medskip
A feature of the propagating kink waves that is of particular interest is to estimate is the observed damping of these waves. The damping can be measured through analysis of the power ratio of the outward to inward propagating waves, ${P(f)}_{ratio}$. As mentioned, this has been performed previously by a number of authors for a single case-study \citep{verth2010a,VERetal2013b,PASetal2015,Tiwari_2019,Montes_Solis_2020}.

The velocity power as a function of frequency for the inward, $P_{in}(f)$, and outward $P_{out}(f)$, component of the waves are obtained by summing the velocity power spectra in the $k$-direction. For each loop, the inward and outward power spectra are averaged over the neighbouring wave paths to suppress the variability in the power spectra. From this one-dimensional averaged wave power, the ratio of the outward to inward power, $\langle{P(f)}\rangle_{ratio}$, is determined by taking the ratio of the power at corresponding positive and negative frequencies. 

Following \cite{verth2010a}, the function to model the ratio of the power spectra is defined as follows:
 \begin{equation}
\langle{P(f)}\rangle_{ratio}=\frac{P_{out}}{P_{in}} \exp\left(\frac{2L}{\upsilon_{ph}\xi}f\right), \label{eqn:power_rat}
\end{equation} 
where $L$ is the {{half-}}loop length, $v_{ph}$ is the propagation speed and $\xi$ is the equilibrium parameter (or quality factor) that provides a measure of the strength of the wave damping. The factor $P_{out}/P_{in}$ can be interpreted as the power ratio at the loop footpoint.

Estimates for $\xi$ are obtained by fitting the model power ratio given by Equation~\ref{eqn:power_rat} to the data, using a maximum likelihood approach. The associated confidence intervals on the model parameters were estimated by utilising the \textit{Fisher Information}. For a detailed discussion on the statistics of the power ratio and the maximum likelihood approach, see \citeauthor{Tiwari_2019} (\citeyear{Tiwari_2019}, their Section 3.4).


\section{Results and Discussion}{\label{result}}

In total propagating kink waves in 120 individual quiescent loops observed with CoMP are analysed. For each loop, estimates for the loop length, the propagation speed, the power ratio at the loop footpoint ($P_{out}/P_{in}$) and the equilibrium parameter ($\xi$) are obtained. The various parameters that were obtained are listed in Table~\ref{table01}. In the following subsections, a summary of the main properties of the propagating kink waves is provided.

\begin{figure*}[!t]
\centering
\includegraphics[width=1.0\textwidth] {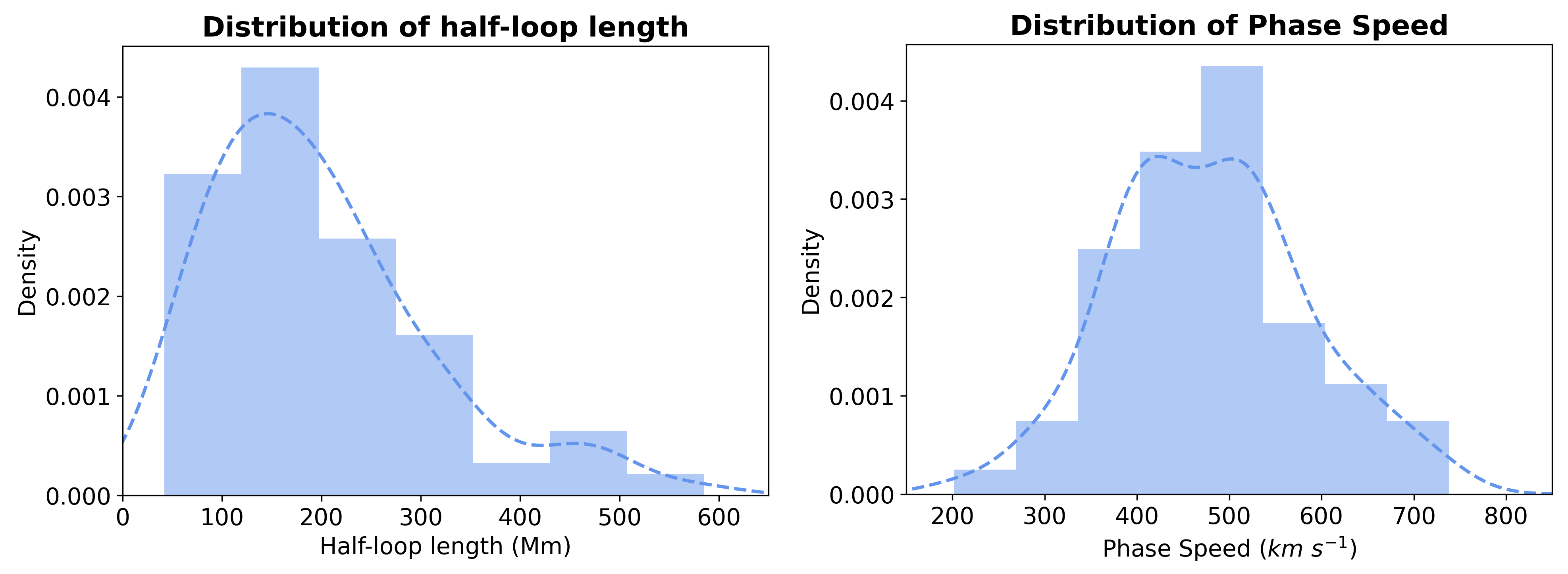}
\caption{\textit{Left panel} shows the distribution of measured half-loop lengths. \textit{Right panel} corresponds to the measured phase speeds for the observed propagating waves. The blue bars and line represent the histogram and Kernel Density Estimate (KDE) estimates for the distributions, respectively.}
\label{figlooplen_dist}
\end{figure*}

\subsection{Loop lengths}

First, a summary of the typical wave-path lengths is provided, which is used as a proxy for the loop length. In the \textit{left panel} of Figure~\ref{figlooplen_dist}, the distribution of the half-loop lengths is shown. The half-loop lengths for the traced coronal loops are in the range of 50-600 Mm. The distribution peaks at around 150-200~Mm, and most of the loops observed are between 50-250 Mm. The number of longer loops is low, as it becomes increasingly difficult to trace the longer loops due to the limited field-of-view of the CoMP instrument. As well as the mentioned issue with visibility of the lower portions of the loops, the observed loop lengths suffer from projection effects, although it is hoped that the selection criteria for the loops minimise this (see Section~\ref{pfss}). Due to the reasons mentioned above, in the case of propagating waves, a looplength always means half the looplength.

\subsection{Propagation speeds}
The distribution of propagation speeds is shown in the \textit{right panel} of Figure~\ref{figlooplen_dist}, with the measured speeds distributed between 200-800~km~{s}$^{-1}$ and peaking around 400-600~km~{s}$^{-1}$. This is consistent with the various propagation speeds reported in the literature \citep{tom2007, tom2009, Liu_2014, MORetal2015, Tiwari_2019, Yang2020b, Yang2020a}. The propagation speed values obtained are averaged over the outward and inward wave propagation speeds. There is some evidence that the outward and inward velocities are different, which can be explained by the presence of flows along the coronal loops. However, the methodology for the measurement of the wave propagation speed is currently not sensitive enough to quantify this, apart from in extreme cases \citep[e.g., in coronal holes, see][]{MORetal2015}. The presence of flows leads to modification of the resonant damping of the kink waves, as described by \citet{SOLetal2011}; consequently, this would require a change in the model for the power ratio that has been used. However, the influence of flows is neglected until they can be inferred more readily.  

\begin{figure*}[!t]
\centering
\includegraphics[width=0.95\textwidth] {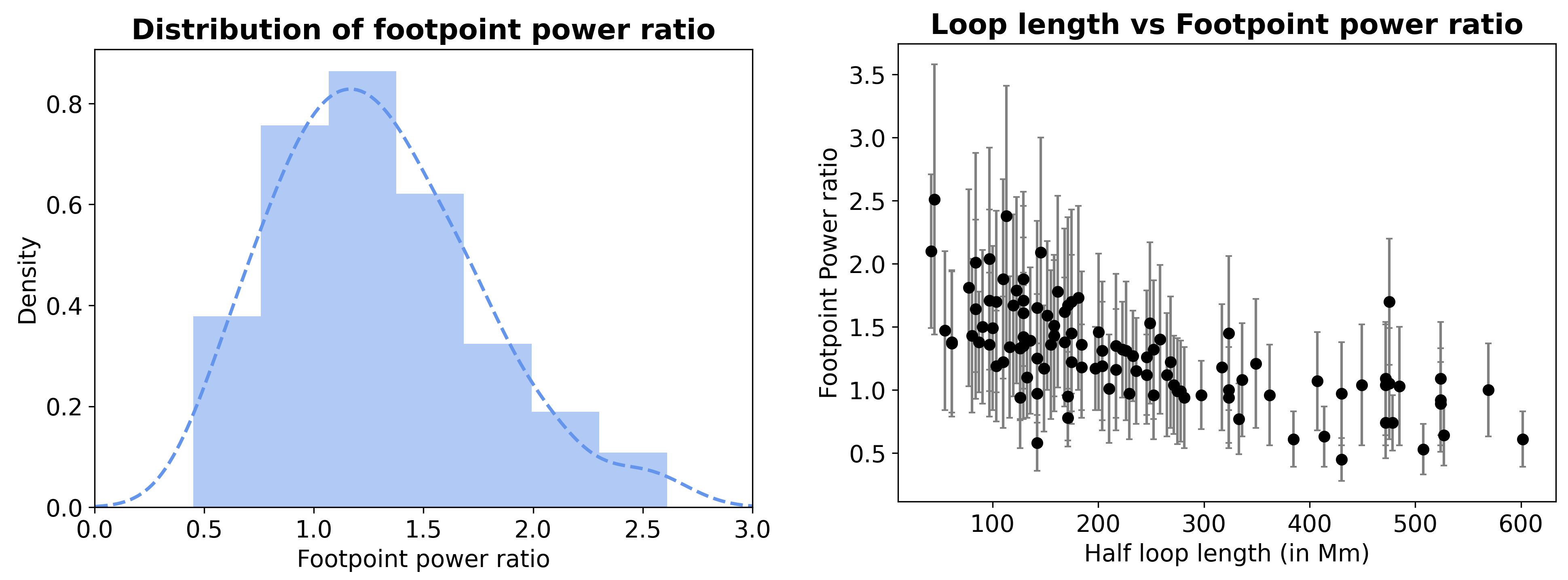}
\caption{\textit{Left panel} shows the distribution of measured footpoint power ratios after performing fitting for the observed waves. The blue bars and line represent the distribution and Kernel Density Estimate (KDE) for the footpoint power ratio for propagating kink waves. \textit{Right panel} highlights the variation of footpoint power ratio with respect to the half-loop length. The black circles represent the footpoint power ratio, and the corresponding error bars are shown in grey.}
\label{figpowrat_dist}
\end{figure*}

\subsection{Power ratio}
The power ratio factor, $P_{out}/P_{in}$, defined in Equation~\ref{eqn:power_rat} is essentially a measure of the power of the waves entering the corona at each footpoint of the loops. The distribution of the estimated values of power ratio is shown in the \textit{left panel} of Figure~\ref{figpowrat_dist}, and has a mean value of $1.29\pm0.04$. While the footpoint power ratio does not provide any information about the driving mechanism, it can be used as a proxy for measuring the energy input at each footpoint of the loop. The driving mechanism of propagating kink waves are thought to be one that acts globally due to the ubiquitous nature of these waves \citep{MORetal2019}. Hence, one would expect that the energy entering the corona through each footpoint will be approximately equal, unless each set of footpoints is located in regions with dissimilar magnetic field strengths. The mean value of the footpoint power ratio supports this hypothesis and is in agreement with the results from previous studies \citep{verth2010a, Tiwari_2019}. The scatter around the value of $1$ could indicate that in some regions of the atmosphere, the driver is weaker/stronger than in others.

However, examination of the behaviour of the power ratio as a function of the length of the coronal loop reveals the footpoint power ratio exhibits a decreasing trend with loop length (\textit{right panel} of Figure~\ref{figpowrat_dist}). The power ratio starts at values close to 1.5 for the shorter loops and tends towards one as the loop length increases. A potential explanation for this trend could be that, for shorter loops, the wave power injected at each footpoint is different, possibly due to different excitation mechanism or due to differences in the frequency/strength of the driver. Why this should be the case for shorter loops only is not evident. 

Instead, it is suggested that the enhanced power ratio for shorter loops is an artefact of the analysis method. If the spatial wavelength of the oscillation is on the order of, or greater than, the length of the wave path used in the analysis, then there can be a leakage of power into negative/positive values of $k$. As an example, the wavelength of the kink modes is $v_{ph} P$, where $v_{ph}$ is the phase speed, and $P$ is the period. Using the values associated with the observed waves, a period of 300~s and phase speed in the range 200-600~km~s$^{-1}$ then gives wavelengths in the range 60 - 180~Mm. Hence, the wavelengths are comparable to the lengths of the shorter loops selected here. The leakage of power from one quadrant of the power spectrum to another might be able to explain the observed deviation from 1. In future work, analysis can be modified to negate the impact of this power leakage, which can be avoided by fitting the power spectrum in $\omega$-$k$ space instead of just frequency.

\subsection{Equilibrium parameter}
\begin{figure*}[!t]
\centering
\includegraphics[width=0.95\textwidth] {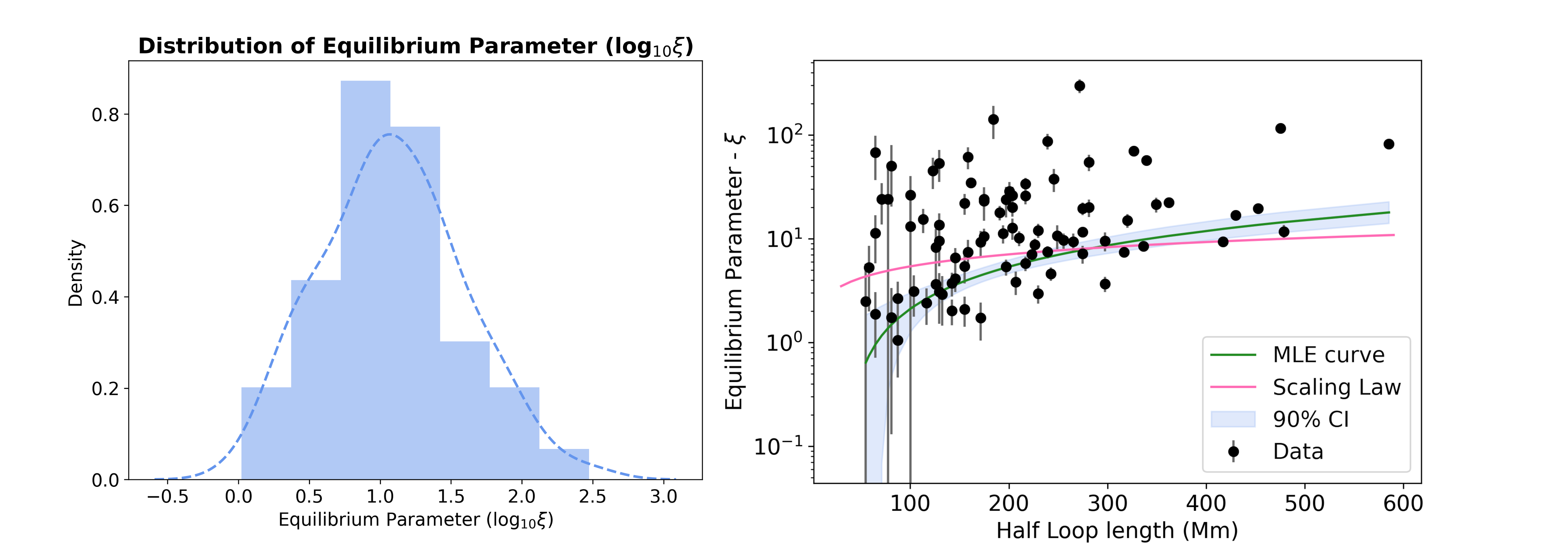}
\caption{\textit{Left panel} shows the distribution of estimated equilibrium parameter $\xi$. The blue bars and line represent the estimates for the distribution from histogram and Kernel Density Estimate methods, respectively. \textit{Right panel} shows the dependence of the equilibrium parameter $\xi$ on the half loop length. The green curve shows the maximum likelihood estimate for a quadratic function of $L$, with the associated confidence intervals. The pink line is the potential variation in $\xi$ from coronal loop scaling laws.}
\label{figdamplen_dist}
\end{figure*}

Perhaps the most interesting parameter estimated here is the equilibrium parameter or quality factor ($\xi$), which quantifies the damping rate of the waves. The distribution of $\xi$ presented in the \textit{left panel} of Figure~\ref{figdamplen_dist} for the positive values. The values of $\xi$ are between 0.89 and $\sim298$; hence they occupy a wide range of values \citep[relative to the standing modes, see][]{Morton_2021}. The distribution illustrates that $\sim80$~\% of the positive $\xi$ observations fall in the range of (0.89,30). Hence, the propagating kink waves can be strongly damped or very weakly damped. The median value is $\sim11$, and a mean value of $\sim18$, which is greater than that found for the standing kink modes \citep[see][for a full discussion of the importance of the observed values of the equilibrium parameter]{Morton_2021}. 

It should also be noted that of the 108 loops identified and studied, 31 of them show signs of power amplification, with a negative value of $\xi$. These are observed only for short loops (less than the half-loop length of 350 Mm). At present, it is unclear whether these results are physical. As already mentioned, short loops typically contain lower signal-to-noise velocity time-series due to their proximity to the occulting disk. However, it has been shown by \cite{SOLetal2011} that flows can play an important role and can lead to amplification of waves. The amplification of waves is always in competition with wave damping mechanisms. However, as discussed, the values of $\xi$ are typically large, which implies a weak damping. Hence, in loops with weak damping, there is an improved chance to observe any amplification that may arise from flows or other mechanisms. These shorter loops with negative $\xi$ need further investigation. 

\subsection{Equilibrium parameters relationship with {{half-}}loop length}

\citet{Tiwari_2019} gave evidence in favour of dependence between the equilibrium parameter ($\xi$) and the loop length, although there were only seven loops analysed in that study. The additional measurements made here enables us to examine this dependence further. As mentioned, the loops selected for this study correspond to loops being orientated such that the longitudinal axes are predominantly in the plane-of-sky. It implies that the longer loops reach higher altitudes in the corona. The \textit{right panel} of Figure~\ref{figdamplen_dist} displays a scatter plot, revealing a range of equilibrium parameters are possible for all loop lengths. The equilibrium parameter also shows a distinct behaviour with increasing loop length. As indicated by the results in \citet{Tiwari_2019}, as loop length increases, there is an increase in the value of the damping equilibrium parameter. 

In order to show this relationship, whether several simple models of the form $\hat{\xi}=f(L)$ could describe the data is examined. The models examined were: constant, linear, quadratic, square root and log. Each model also contained a constant term. The models were fit to the data assuming that a Normal distribution describes the likelihood of the data; hence the negative log-likelihood of the form is minimised:
$$
-2\ln \mathcal{L}=\sum_{i=1}^{N}\frac{(\xi_{i}-\hat{\xi(L)})^2}{\sigma_i^2},
$$
where $\xi_{i}$ is the observed value, $\hat{\xi}(L)$ is the model prediction, and $\sigma_i$ is the uncertainty on $\xi_i$. The out-of-sample prediction error is estimated to test the ability of each model to describe the data. First,leave-one-out cross-validation is utilised, using the mean value of the negative log-likelihood as a measure of the test error. It turns out all models (except the constant) show a similar ability to describe the data, with the quadratic and linear models performing the best, although the difference between the non-constant models is small. Moreover, the Akaike Information Criteria calculation supports the results from cross-validation and confirms that the constant model performs the worst. 

In the \textit{right panel} of Figure~\ref{figdamplen_dist}, the results from the quadratic model is displayed. The uncertainty on the model curve is calculated by performing 10,000 bootstrap simulations of the fitting procedure, then using the percentile method to estimate the point-wise confidence intervals. The increase in $\xi$ with loop length is evident. It is noted that the data also appear to show there is a lower bound to the possible values of $\xi$ for the loops that increases with loop length. 

\medskip

The implication of the increase in $\xi$ with loop length is that the propagating kink waves are subject to a reduced rate of damping for longer loops. As discussed in \cite{Tiwari_2019}, a physical explanation for the apparent decrease in damping rates can be made. Assuming that resonant absorption is the mechanism acting to provide the observed frequency-dependent damping, the quality factor for kink modes is given by 
\begin{equation}
\xi = \alpha \frac{R}{l}\frac{\zeta+1}{\zeta-1}
\label{eq:equib}
\end{equation}
where $R$ is loop radius, $l$ is the thickness of the density inhomogeneity layer, $\zeta=\rho_i/\rho_e$ is the ratio of the internal and external densities of the magnetic flux tube, respectively, and $\alpha$ is a constant whose value describes the gradient in density across the resonant layer. It is suggested that the key factor in understanding the observed behaviour would be the density ratio between the internal and external plasma. Quiescent coronal loops should be subject to similar heating rates; hence the rate of associated chromospheric evaporation is similar. If this is true, then the average density of the longer loops is likely to be less than those of shorter loops. This will lead to the density ratio ($\rho_i/\rho_e$) for longer loops being, on average, smaller than for the shorter loops compared to the ambient plasma.

As a basic examination of this premise, the scaling laws for dynamic loops derived in \cite{Bradshaw_2020} is utilised. It is noted that the scaling laws are derived under the assumption of constant pressure along the loop, which may limit the applicability of the results to short, hot loops.
From their Eq.~(45), the loop apex density, $n_m$, is related to the heating rate, $E_H$, by
$$
n_m = \left[ c(M)E_H L^{1/4} \right]^{4/7} ,
$$
where $L$ is the half-loop length, and $c(M)$ is a function of the Mach number. This expression is re-written in terms of the energy flux, $F_H=E_HL$, giving,
\begin{equation}
n_m
= \left[ c(M)F_H L^{-3/4} \right]^{4/7}.
\end{equation}
The loop number density can be thought of as a combination of an initial number density ($n_0$) (due to some basal heating, $E_{H_0}$), which is assumed to be equal to ambient plasma ($n_e$), plus an additional density, $n_1$, from the evaporation of the chromospheric/transition region due to some heating event ($E_{H_{1}}$) associated with the loop, i.e.
$$
n_m = n_0(E_{H_0})+ n_1(E_{H_{1}}).
$$
The density ratio $\zeta$ can be defined as:
\begin{eqnarray}
\zeta &=&\frac{n_m}{n_e}\\
&=& 1+ \frac{F_{H_1}^{4/7}}{(2.4\times10^{-15})^{4/7}Mn_e}L^{-3/7}
\label{eq:zeta}
\end{eqnarray}
This expression suggests that the density ratio may depend upon the length of the coronal loop, with the over-density of the loop decreasing as the loop length increases. 

Substituting the expression for $\zeta$ into Eq.~\ref{eq:equib} to estimate the quality factor as a function of loop length. The values of $\xi$ by providing some reasonable values for the quiet Sun parameters are calculated, assuming: the energy flux is $F_H\sim$200 W m$^{-2}$ = 2$\times10^5$ ergs s$^{-1}$ cm$^{-2}$; the electron density is $n_e=10^{8.5}$ cm$^{-3}$; the Mach number is $0.1$ (corresponding to flows of $\sim10$~km/s); $\alpha=2/\pi$; and $l/R=1$. The dependence of the quality factor on loop length from the scaling law theory is shown in Figure~\ref{figdamplen_dist}. While it does not match the curve from the model fitting, the results support our physical explanation for an increase in the equilibrium parameter as a function of loop length.

\medskip

\begin{figure}[!h]
\centering
\includegraphics[scale=0.45] {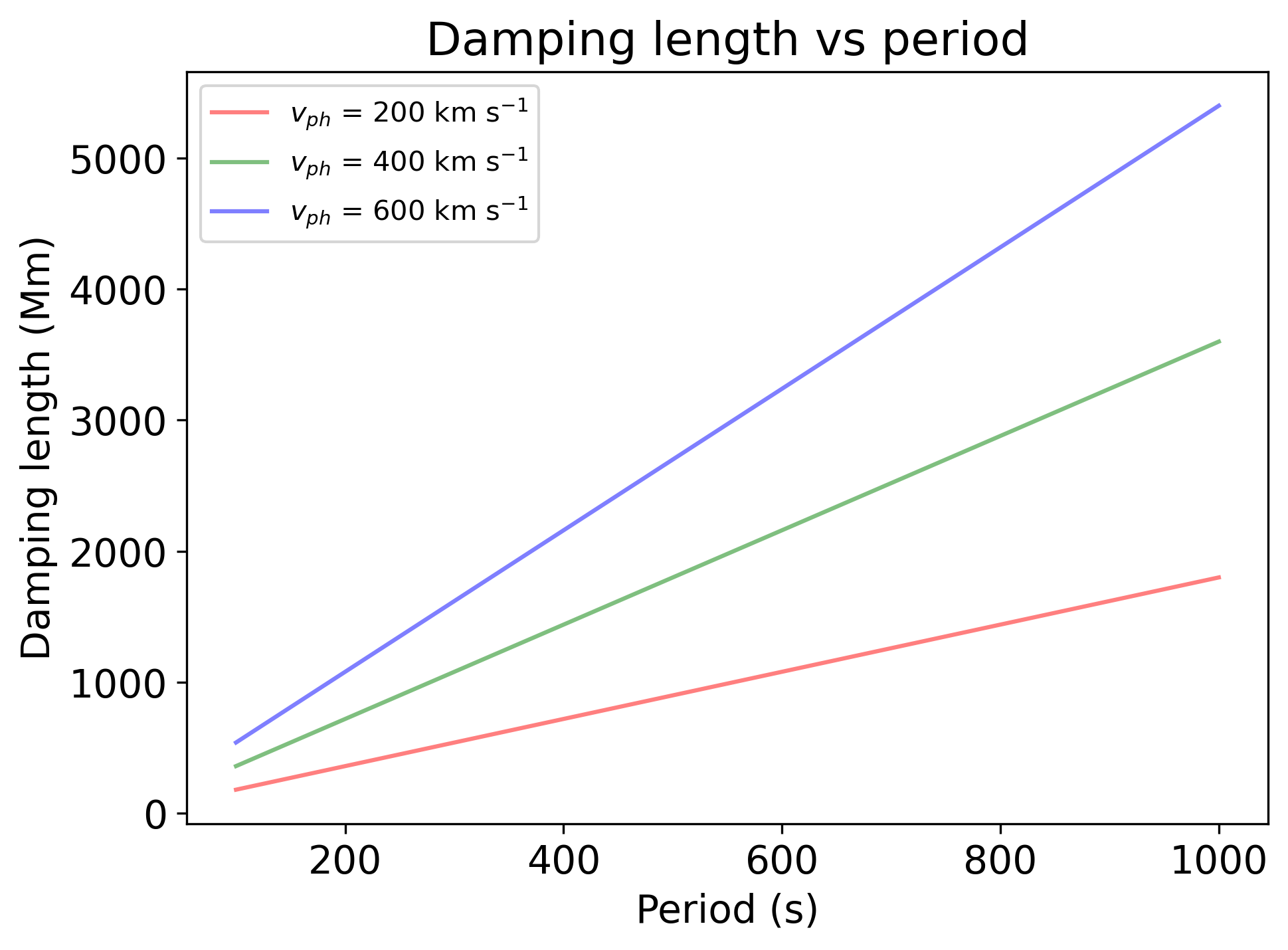}
\caption{Estimated variation of damping length with phase speed, calculated using typical values for period and the median value of the equilibrium parameter.}
\label{figdamplen_var}
\end{figure}

It is always worthwhile drawing comparisons between related results. To our knowledge, there is only one previous estimate for the damping rate
of Alfv\'enic modes in the quiescent corona \citep[excluding][]{verth2010a, Tiwari_2019}, which was estimated in \citet{hahn2014}\footnote{In \citet{hahn2014}, measurements of
non-thermal line widths from Hinode/EIS data are assumed to represent Alfv\'enic waves. It is likely that the non-thermal broadening is due to the under-resolved kink waves, e.g. \cite{MCIDEP2012}, \cite{Pant_2019_dark}.}. 
They provide estimates for the damping lengths of the waves, finding a broad distribution (up to 500~Mm) with a median value between 100-200~Mm. In order to provide a comparison to their results, the estimated quality factors should be converted to damping lengths. The damping length ($L_{d}$) can be calculated using
$$
L_{d}=\xi \lambda_{prop},
$$ 
where $\lambda_{prop}=v_{ph} P$. Substituting typical values of the period ($P$) of the waves observed by {\comp} (100 - 1000~s) and the measured phase speeds (see Table~\ref{table01}) in this expression given the damping lengths. Figure~\ref{figdamplen_var} shows the estimated damping lengths as a function of the period for three different values of $v_{ph}$, using the median value of $\xi$ from measurements in this study. The damping lengths reported in \cite{hahn2014} are comparable to the estimated damping lengths for the shorter period waves from the CoMP observations (Figure~\ref{figdamplen_var}). The damping lengths given by \citet{hahn2014} will themselves, of course, be related to waves with a particular range of periods. However, it is not straightforward to ascertain the periods of the waves encapsulated in the non-thermal line-widths.

A value of $\xi$ from the information in \citet{hahn2014} can also be estimated. Although the assumed values of propagation speed, $v_{ph}$, are not given in \citet{hahn2014}, by inverting the given equation for the energy flux, $F$, namely,
$$
F=\rho \langle \delta v^2 \rangle v_{ph},
$$
where $\langle \delta v \rangle$ is the velocity amplitude from the non-thermal widths, gives the propagation speed. The given values of energy flux ($5.5 \times 10^{5}$~ergs~s$^{-1}$~cm$^{-2}$), non-thermal widths (30~km~s$^{-1}$) and electron density ($5 \times 10^{8}$~cm$^{-3}$) is also utilised to find $v_{ph}$ = 550~km~s$^{-1}$. Hence, using the damping lengths of 100-200~Mm, the quantity $\xi P$ = 180 - 360~s. The Hinode/EIS data used in their study is integrated over 60~s. Assuming that only waves with periods less than 60~s contribute to the line broadening (which is a very conservative assumption), then $\xi$ = 3-6. These values are likely overestimates for their study but are broadly in agreement with the range of values found in this study (e.g. Figure~\ref{figdamplen_dist}).

\subsection{Comparison with observations of different modes of kink waves} \label{standing_waves}

Before concluding, a light-touch comparison between the properties of the propagating kink waves observed here and the previous studies of the two standing kink modes, i.e. damped and decay-less, is also provided. The comparison is worthwhile to highlight that the standing and propagating modes are found in coronal loops with significantly different plasma conditions.  

\smallskip

For the damped standing kink waves, the 
catalogue of events compiled by \citet{goddard2016} and \citet{Nechaeva_catalogue}, using data from SDO/AIA is used. All the observations from these data sets which did not have any associated period or damping time information are removed. The total number of cases after this selection is 103 events over the course of solar cycle 23.

In the catalogue of standing kink waves, the loop length is estimated under the assumption that the loops are close to the semi-circular shape by either measuring the projected distance between the footpoints or by the apparent height. For each oscillation, the amplitude of the initial displacement and initial oscillation amplitude was given, along with the period. The mode of the standing kink waves here is assumed to be the fundamental. 

The kink speed, which is not provided in the catalogue of \citet{Nechaeva_catalogue}, is calculated as follows:
\begin{equation} 
 c_{kink} = \frac{\omega}{k} = \frac{4 L}{P},
 \label{kink_speed}
\end{equation}
where $c_{kink}$ is the kink speed, $L$ is the half loop length, and $P$ is the period of the waves observed. Furthermore, the equilibrium parameter is also calculated by using the measured damping time of the oscillations and the periods, i.e. $\xi=\tau/P$.

\smallskip
For the decay-less kink waves, the catalogue put together by \citet[see their table A.1]{Anfino2015}\footnote{The table was scraped using Beautifulsoup module \citep{BS4}.} is used. 
The catalogue provides a study of 71 observations of decay-less kink waves in 21 active regions (NOAA 11637–11657) between December 2012–January 2013. 

\begin{table*}[!th]
\begin{longtable*}[c]{l c c c}
\hline

Properties &
 Damped Standing &
 Damped Propagating &
 Decayless Standing  \\ 
\hline

Loop length (Mm)         & 324  & 394  & 219  \\ 
Period (s)          & 7.3  &  -    & 4.3  \\ 
Amplitude (Mm)         & 5    &  -    & 0.17 \\ 
Equilibrium Parameter & 1.8  & 11.4 &    -  \\ 
Kink Speed (km s$^{-1}$)        & 1336 & 482  & 1743 \\ 
\hline
\caption{Median values of various characteristics of the different observed kink modes in the solar corona.}
\label{tab:summary}\\
\end{longtable*}
\end{table*}

From both of the catalogues, the median values for the loop length, period, amplitude and kink speed (see Table~\ref{tab:summary}) is shown \footnote{The loop length in case of the standing and the decayless kink waves are the full semi-circular loop lengths as seen in AIA/SDO, while in the case of the propagating kink waves the loop lengths refer to the half-loop length as seen in the \comp{} FOV}. For the damped waves, the median value of the equilibrium parameter is also presented.

\smallskip
In terms of the length of coronal loops which support the oscillations, the loops studied in this paper here have a similar distribution to those from the damped and decay-less standing wave studies. The propagating kink wave catalogue also contains several longer loops. The reader is reminded that the loop lengths measured by \comp \ and SDO/AIA are not directly comparable. The \comp \ instrument has an occulting disk that obscures the corona below 1.05 $R_{\ Sun}$. Hence the measurements for the coronal loops' length do not begin near the footpoints, and it is likely an underestimation of the length of the CoMP loops (a rough correction being the addition of 70~Mm to the given values). On the other hand, SDO/AIA images show the solar-disk, often making the coronal footpoints of the loop visible in the images. 

\smallskip

One of the main difference between the two loop populations is the measured kink speed. In the case of standing waves, the estimated kink speeds have medians of $\sim$1300~km~{s}$^{-1}$ and $\sim$1700~km~{s}$^{-1}$ for damped and decay-less, respectively. While for the propagating waves, the median value of propagation speed is substantially smaller at $\sim$480~km~{s}$^{-1}$. The contrast of these values reflects the sizeable differences in the magnetic field strengths between the regions where these waves are observed and somewhat the different densities. The  
electron density in the quiet Sun ($10^{8-9}$~cm$^{-3}$) is less than that of active regions 
($10^{9-10}$~cm$^{-3}$), which indicates the magnetic field strength must be substantially weaker in the quiescent Sun. This is borne out by estimates of the magnetic field strength, which in the active region coronal loops lies in the range of 4-30~Gauss \citep[e.g.][]{naka2001, white2012}\footnote{It should be noted that the magnetic field 
measurements obtained by coronal magneto-siesmology provide an underestimate of the magnetic field values \citep{verwichte2013a}.}, while the magnetic field of the quiet Sun loops are estimated to 
be between 1-9 Gauss \citep{MORetal2015,comp_mag,Yang2020a, Yang2020b}.   

\smallskip

The difference in plasma parameters will ultimately bring about differences in how the waves/oscillations evolve as a function of time and/or distance. In fact, in the companion paper, \citet{Morton_2021}, a comprehensive discussion on the implications of differences in the equilibrium parameter found for the standing waves ($\xi_{median} =1.8$) and the propagating waves ($\xi_{median} =11.4$) is provided.

\section{Summary and Conclusion}\label{conclusion}
The details of a catalogue of quiescent coronal loops observed with the CoMP instrument are provided, all of which show evidence for the presence of propagating kink waves. The catalogue is used to undertake a statistical study of the propagating kink waves providing estimates for the damping rate and propagation speeds of the waves,
presenting some details of how the propagating kink waves evolve. It is found that the equilibrium parameter, which quantifies the degree of wave damping,
has a broad range of values, which indicates that in some of the coronal loops, the propagating kink waves are only weakly damped \cite[this is aspect is discussed further in][]{Morton_2021}.
The damping length of the propagating kink waves is also estimated, which is found to be broadly comparable to the previous estimates of \citep{hahn2014}. Moreover, the study also finds that there is a relationship between the degree of damping and loop length, with waves propagating along longer loops typically experiencing reduced damping, verifying claims of \citet{Tiwari_2019}. The suggested reason for this
behaviour is related to longer loops having a lower average density contrast, potentially due to limits on the amount of mass that can be evaporated during heating events associated with a given heating rate.

The study also reports the amplification of waves, the source of which is unclear at this time. 

\smallskip

A brief comparison of the observed properties of the propagating waves to the standing modes is also presented.
Notable differences between propagation speed and damping rates are found, with the contrast being due to the dissimilar plasma and magnetic environments of the two populations of loops that support the waves. The standing kink waves have been reported predominantly in loops with at least one footpoint in an active region; however, the propagating kink waves have been reported to be ubiquitous in the solar corona. 

\smallskip

It is envisaged that the catalogue {{of propagating kink waves}} will provide the community with the foundation for further study of propagating kink waves in the quiet solar corona\footnote{The catalogue in Table~\ref{tab:summary} is available as a csv file.}. Many potential studies can exploit the propagating kink waves to further probe the plasma conditions in the quiescent loops, with the potential to incorporate density measurements from the Fe XIII line pair that CoMP also observes and provide estimates of magnetic field and flows through magneto-seismology \citep{MORetal2015, Yang2020a}. This will ultimately enable us to develop a clear picture of how the propagating kink waves evolve in the quiescent corona and determine their role in plasma heating.
Moreover, it is emphasised that there is a need for 3D MHD simulations of kink wave propagation in quiescent coronal loops to aid our understanding of the role of resonant absorption in the damping of propagating kink waves. 

\begin{acknowledgments}

We would like to acknowledge Tom Van Doorsselaere, Norbert Magyar, Marcel Goossens and Stephen Bradshaw for valuable discussions.
A.K.T is supported by the European Union's Horizon 2020
research and innovation programme under grant agreement No 824064 (\href{https://projectescape.eu/}{ESCAPE}). R.J.M. is supported by a UKRI Future Leader Fellowship (RiPSAW - MR/T019891/1). J.A.M. is supported by the Science and Technology Facilities Council (STFC) via grant number ST/T000384/1. The authors acknowledge the work of the National Center for Atmospheric Research/High Altitude Observatory CoMP instrument team. The authors also acknowledge STFC via grant number ST/L006243/1 and for IDL support.
\end{acknowledgments}

\pagebreak
\appendix
\startlongtable
\begin{deluxetable*}{|c|c|c|c|c|c|c|c|}
\tablecaption{Measured loop parameters and wave parameters obtained from observations.\label{table01}}
\tablehead{
\colhead{Loop no.} & 
\colhead{Solar-X} & 
\colhead{Solar-Y}& 
\colhead{Date} &
\colhead{Half Loop Length} &
\colhead{$\xi$} &
\colhead{Power Ratio} &  
\colhead{Phase speed}  \\ 
\colhead{} & 
\colhead{(arcsec)} &
\colhead{(arcsec)}  & 
\colhead{} &  
\colhead{(in Mm)} &
\colhead{} &
\colhead{} &
\colhead{(km s$^{-1}$)} 
} 
\startdata
       1 &       798.50 &      -819.50 & 20120423 &           42 &   -1.53 $\pm$ 1.27 &  2.1 $\pm$ 0.61 &    291 $\pm$ 2 \\ \hline
        2 &     -1026.30 &      -238.20 & 20120514 &           42 &   -8.04 $\pm$ 5.78 & 2.17 $\pm$ 0.72 &   415 $\pm$ 15 \\ \hline
        3 &       980.56 &       422.56 & 20120626 &           45 &   -0.95 $\pm$ 2.07 & 2.51 $\pm$ 1.07 &   561 $\pm$ 11 \\ \hline
        4 &      -928.26 &      -549.67 & 20120626 &           45 &    -1.89 $\pm$ 2.0 & 2.51 $\pm$ 1.07 &   561 $\pm$ 11 \\ \hline
        5 &     -1136.86 &       -57.01 & 20120410 &           55 &    2.49 $\pm$ 3.23 & 1.47 $\pm$ 0.63 &    373 $\pm$ 4 \\ \hline
        6 &     -1045.91 &      -251.79 & 20120514 &           58 &   -3.03 $\pm$ 1.64 &  2.0 $\pm$ 0.65 &   412 $\pm$ 15 \\ \hline
        7 &       504.41 &       944.64 & 20120410 &           58 &    5.28 $\pm$ 3.29 & 1.47 $\pm$ 0.63 &    373 $\pm$ 4 \\ \hline
        8 &       972.59 &       356.58 & 20120421 &           65 &   11.35 $\pm$ 5.55 & 0.86 $\pm$ 0.25 &   481 $\pm$ 11 \\ \hline
        9 &       911.23 &       553.31 & 20120719 &           65 &    1.88 $\pm$ 1.17 & 1.37 $\pm$ 0.58 &    402 $\pm$ 5 \\ \hline
       10 &     -1057.90 &        96.35 & 20120810 &           65 &  67.62 $\pm$ 30.86 & 1.38 $\pm$ 0.56 &    343 $\pm$ 6 \\ \hline
       11 &      1049.77 &      -168.19 & 20120121 &           71 &  24.04 $\pm$ 10.33 & 0.89 $\pm$ 0.33 &    376 $\pm$ 6 \\ \hline
       12 &     -1162.06 &       -39.82 & 20120410 &           78 &  24.06 $\pm$ 29.71 & 1.81 $\pm$ 0.78 &    512 $\pm$ 8 \\ \hline
       13 &       549.04 &      -918.78 & 20120410 &           81 &  50.13 $\pm$ 29.75 & 1.81 $\pm$ 0.78 &    512 $\pm$ 8 \\ \hline
       14 &      1098.97 &       124.52 & 20120626 &           81 &    1.74 $\pm$ 1.61 & 1.43 $\pm$ 0.61 &    389 $\pm$ 8 \\ \hline
       15 &      -883.66 &      -646.84 & 20120410 &           84 &   -3.74 $\pm$ 2.96 & 1.64 $\pm$ 0.71 &    365 $\pm$ 4 \\ \hline
       16 &       534.14 &       943.70 & 20120410 &           87 &   -1.87 $\pm$ 1.23 & 2.01 $\pm$ 0.87 &   566 $\pm$ 10 \\ \hline
       17 &     -1046.14 &       -52.13 & 20120410 &           87 &    -7.76 $\pm$ 3.1 & 1.64 $\pm$ 0.71 &    365 $\pm$ 4 \\ \hline
       18 &     -1052.56 &       214.08 & 20120423 &           87 &    2.65 $\pm$ 1.21 &  1.38 $\pm$ 0.4 &    300 $\pm$ 7 \\ \hline
       19 &      -343.39 &      1002.95 & 20120121 &           87 &    1.05 $\pm$ 0.59 & 0.58 $\pm$ 0.43 &   526 $\pm$ 10 \\ \hline
       20 &     -1055.64 &       -13.30 & 20120810 &           91 &  -17.92 $\pm$ 5.32 &  1.5 $\pm$ 0.61 &    308 $\pm$ 4 \\ \hline
       21 &       423.22 &     -1074.57 & 20130717 &           97 &    -2.0 $\pm$ 2.37 & 2.04 $\pm$ 0.88 &   630 $\pm$ 36 \\ \hline
       22 &      -628.86 &      -842.94 & 20120719 &           97 & -27.77 $\pm$ 11.13 & 1.71 $\pm$ 0.72 &    427 $\pm$ 6 \\ \hline
       23 &       925.44 &      -478.88 & 20130717 &          100 &  13.13 $\pm$ 13.96 & 1.49 $\pm$ 0.65 &    557 $\pm$ 9 \\ \hline
       24 &      -472.36 &      -967.83 & 20130717 &          100 &   26.27 $\pm$ 14.0 & 1.49 $\pm$ 0.65 &    557 $\pm$ 9 \\ \hline
       25 &     -1036.89 &      -204.53 & 20130717 &          100 &   -4.13 $\pm$ 2.58 & 2.04 $\pm$ 0.88 &   630 $\pm$ 36 \\ \hline
       26 &      -339.56 &      1007.43 & 20120121 &          103 &     3.1 $\pm$ 1.33 & 1.19 $\pm$ 0.44 &    537 $\pm$ 8 \\ \hline
       27 &      1028.62 &       357.55 & 20120719 &          107 &   -8.54 $\pm$ 2.71 &  1.7 $\pm$ 0.72 &    361 $\pm$ 7 \\ \hline
       28 &     -1052.56 &      -218.54 & 20130717 &          113 &   -6.06 $\pm$ 2.56 & 2.38 $\pm$ 1.03 &    489 $\pm$ 9 \\ \hline
       29 &       693.58 &      -833.45 & 20120410 &          113 &   15.42 $\pm$ 4.05 & 1.22 $\pm$ 0.52 &    315 $\pm$ 4 \\ \hline
       30 &       993.55 &       532.00 & 20120121 &          116 &     2.4 $\pm$ 0.92 & 0.77 $\pm$ 0.28 &   522 $\pm$ 15 \\ \hline
       31 &      1090.09 &        29.57 & 20120810 &          123 &   -7.75 $\pm$ 2.86 & 1.79 $\pm$ 0.74 &   512 $\pm$ 11 \\ \hline
       32 &      -935.88 &       549.79 & 20120410 &          123 &  45.25 $\pm$ 15.16 & 1.67 $\pm$ 0.72 &    439 $\pm$ 7 \\ \hline
       33 &      1111.05 &       -63.70 & 20120121 &          126 &    3.65 $\pm$ 0.53 & 0.45 $\pm$ 0.17 &    202 $\pm$ 5 \\ \hline
       34 &      -808.14 &      -692.31 & 20120410 &          126 &    8.26 $\pm$ 4.19 & 1.33 $\pm$ 0.57 &   675 $\pm$ 15 \\ \hline
       35 &      -647.63 &      -850.12 & 20120719 &          129 &   13.62 $\pm$ 3.91 & 1.33 $\pm$ 0.56 &    404 $\pm$ 5 \\ \hline
       36 &     -1050.43 &        80.15 & 20130717 &          129 &  53.42 $\pm$ 18.21 & 1.71 $\pm$ 0.75 &   461 $\pm$ 11 \\ \hline
       37 &      -983.72 &      -512.77 & 20120423 &          129 &    9.54 $\pm$ 4.12 & 1.42 $\pm$ 0.41 &    440 $\pm$ 8 \\ \hline
       38 &      -418.59 &      1004.38 & 20120121 &          129 &    -3.2 $\pm$ 1.33 &  1.61 $\pm$ 0.6 &   643 $\pm$ 11 \\ \hline
       39 &       737.25 &      -837.03 & 20120410 &          129 &    3.08 $\pm$ 1.57 &  0.94 $\pm$ 0.4 &   677 $\pm$ 15 \\ \hline
       40 &       623.30 &      -865.44 & 20120410 &          133 &  -11.43 $\pm$ 3.38 & 1.35 $\pm$ 0.58 &    415 $\pm$ 6 \\ \hline
       41 &       771.64 &      -726.14 & 20120423 &          133 &    2.91 $\pm$ 1.46 &  1.1 $\pm$ 0.32 &    515 $\pm$ 9 \\ \hline
       42 &     -1048.32 &        40.14 & 20120121 &          133 &  -18.39 $\pm$ 5.91 & 1.88 $\pm$ 0.69 &    537 $\pm$ 8 \\ \hline
       43 &     -1057.14 &      -255.49 & 20120514 &          133 &   -11.62 $\pm$ 2.6 & 1.76 $\pm$ 0.58 &   410 $\pm$ 13 \\ \hline
       44 &      1102.70 &      -207.84 & 20120121 &          142 &    2.03 $\pm$ 0.57 & 0.58 $\pm$ 0.22 &    445 $\pm$ 5 \\ \hline
       45 &      1029.24 &      -114.68 & 20120421 &          142 &    3.73 $\pm$ 0.96 & 0.72 $\pm$ 0.21 &   516 $\pm$ 28 \\ \hline
       46 &       830.82 &       774.90 & 20120810 &          146 &    4.11 $\pm$ 1.04 &  0.97 $\pm$ 0.4 &    402 $\pm$ 9 \\ \hline
       47 &       794.71 &       807.84 & 20120810 &          146 &    6.57 $\pm$ 1.56 & 1.25 $\pm$ 0.51 &    389 $\pm$ 6 \\ \hline
       48 &     -1054.36 &        48.97 & 20130717 &          149 &   -3.31 $\pm$ 0.94 & 2.09 $\pm$ 0.91 &    418 $\pm$ 9 \\ \hline
       49 &      -419.76 &      1012.89 & 20120121 &          155 &    2.09 $\pm$ 0.67 & 0.68 $\pm$ 0.25 &    587 $\pm$ 8 \\ \hline
       50 &      1098.00 &      -220.96 & 20120121 &          155 &   21.99 $\pm$ 5.02 & 1.59 $\pm$ 0.59 &    445 $\pm$ 6 \\ \hline
       51 &       566.20 &      -934.28 & 20120121 &          155 &    5.44 $\pm$ 1.73 &  0.83 $\pm$ 0.3 &   608 $\pm$ 11 \\ \hline
       52 &       628.47 &      -875.99 & 20120410 &          158 &  61.54 $\pm$ 14.87 & 1.36 $\pm$ 0.59 &    409 $\pm$ 6 \\ \hline
       53 &     -1063.02 &        22.18 & 20120120 &          158 &    7.42 $\pm$ 2.36 & 1.36 $\pm$ 0.57 &   549 $\pm$ 15 \\ \hline
       54 &      1116.94 &      -131.81 & 20140102 &          162 &   -25.26 $\pm$ 7.1 &  1.43 $\pm$ 0.6 &   499 $\pm$ 20 \\ \hline
       55 &      1139.13 &       203.94 & 20120121 &          162 &   34.77 $\pm$ 4.23 & 1.51 $\pm$ 0.56 &    246 $\pm$ 3 \\ \hline
       56 &       782.06 &      -736.51 & 20120423 &          171 &    1.73 $\pm$ 0.69 & 0.78 $\pm$ 0.23 &    507 $\pm$ 7 \\ \hline
       57 &      1094.88 &      -209.72 & 20120810 &          171 &   -7.86 $\pm$ 2.34 & 1.62 $\pm$ 0.66 &   577 $\pm$ 12 \\ \hline
       58 &       565.98 &      -942.74 & 20120121 &          171 &    9.31 $\pm$ 2.61 & 1.38 $\pm$ 0.51 &    598 $\pm$ 9 \\ \hline
       59 &      -951.82 &       557.94 & 20120410 &          175 &   -37.47 $\pm$ 7.9 & 1.45 $\pm$ 0.62 &    394 $\pm$ 6 \\ \hline
       60 &       638.63 &      -868.98 & 20120810 &          175 &   23.09 $\pm$ 8.19 & 1.22 $\pm$ 0.49 &   716 $\pm$ 22 \\ \hline
       61 &      1057.52 &       264.13 & 20120121 &          175 &   10.47 $\pm$ 2.08 & 0.95 $\pm$ 0.35 &   433 $\pm$ 10 \\ \hline
       62 &       758.52 &       724.19 & 20120719 &          175 &   24.02 $\pm$ 4.61 &  1.7 $\pm$ 0.73 &    362 $\pm$ 6 \\ \hline
       63 &      -958.90 &       561.96 & 20120410 &          184 & -109.0 $\pm$ 22.22 & 1.36 $\pm$ 0.58 &    405 $\pm$ 7 \\ \hline
       64 &       787.18 &      -744.80 & 20120423 &          184 & -99.76 $\pm$ 32.84 & 1.18 $\pm$ 0.34 &    485 $\pm$ 7 \\ \hline
       65 &      1072.37 &        74.36 & 20120719 &          184 & 141.25 $\pm$ 49.77 & 1.73 $\pm$ 0.73 &   717 $\pm$ 18 \\ \hline
       66 &      1066.31 &       397.09 & 20120421 &          191 &   17.88 $\pm$ 2.85 & 1.02 $\pm$ 0.29 &    473 $\pm$ 9 \\ \hline
       67 &     -1062.69 &         9.00 & 20120121 &          194 &   11.22 $\pm$ 2.23 & 0.74 $\pm$ 0.28 &    475 $\pm$ 8 \\ \hline
       68 &     -1080.66 &       119.18 & 20120423 &          197 &   23.79 $\pm$ 7.75 & 1.17 $\pm$ 0.33 &    511 $\pm$ 8 \\ \hline
       69 &      1034.72 &       361.26 & 20120421 &          197 &    5.36 $\pm$ 0.95 &  0.8 $\pm$ 0.23 &   501 $\pm$ 25 \\ \hline
       70 &     -1086.52 &        24.93 & 20120120 &          197 &  -33.08 $\pm$ 4.29 & 1.88 $\pm$ 0.79 &    285 $\pm$ 5 \\ \hline
       71 &      -974.84 &      -464.22 & 20120120 &          200 &    28.76 $\pm$ 6.5 & 1.21 $\pm$ 0.51 &   507 $\pm$ 17 \\ \hline
       72 &     -1093.16 &       -58.39 & 20120410 &          204 &   20.08 $\pm$ 3.71 & 1.19 $\pm$ 0.51 &    400 $\pm$ 7 \\ \hline
       73 &      -369.52 &      1028.94 & 20120121 &          204 &    12.7 $\pm$ 2.98 &  1.0 $\pm$ 0.37 &    589 $\pm$ 9 \\ \hline
       74 &     -1063.58 &       117.42 & 20130717 &          204 &   26.12 $\pm$ 5.55 & 1.46 $\pm$ 0.62 &   461 $\pm$ 12 \\ \hline
       75 &       564.80 &      -953.78 & 20120121 &          207 &    3.84 $\pm$ 0.98 & 0.64 $\pm$ 0.24 &   641 $\pm$ 13 \\ \hline
       76 &      -831.80 &      -722.20 & 20120410 &          210 &   10.15 $\pm$ 1.68 & 1.01 $\pm$ 0.43 &    365 $\pm$ 5 \\ \hline
       77 &      1127.22 &      -127.23 & 20120121 &          217 &   33.86 $\pm$ 4.64 & 1.07 $\pm$ 0.39 &    381 $\pm$ 4 \\ \hline
       78 &      -643.34 &      -879.80 & 20120719 &          217 &   25.83 $\pm$ 4.47 & 1.16 $\pm$ 0.48 &    414 $\pm$ 5 \\ \hline
       79 &      1058.06 &       359.66 & 20120421 &          217 &    5.81 $\pm$ 0.93 & 0.76 $\pm$ 0.22 &   501 $\pm$ 25 \\ \hline
       80 &       792.63 &      -769.79 & 20120423 &          223 &  -14.75 $\pm$ 3.96 & 1.32 $\pm$ 0.38 &    483 $\pm$ 8 \\ \hline
       81 &      1101.27 &       268.27 & 20120121 &          223 &    7.05 $\pm$ 0.78 & 0.61 $\pm$ 0.22 &    297 $\pm$ 6 \\ \hline
       82 &      1097.27 &       317.64 & 20120121 &          226 &    8.73 $\pm$ 1.28 & 0.73 $\pm$ 0.27 &    404 $\pm$ 7 \\ \hline
       83 &     -1149.07 &       -37.31 & 20120121 &          230 &   12.04 $\pm$ 1.84 & 0.97 $\pm$ 0.36 &    441 $\pm$ 6 \\ \hline
       84 &      1039.45 &       230.53 & 20120121 &          230 &     2.96 $\pm$ 0.6 &  0.53 $\pm$ 0.2 &   523 $\pm$ 15 \\ \hline
       85 &     -1045.10 &        97.27 & 20120423 &          233 &  -162.2 $\pm$ 45.2 & 1.27 $\pm$ 0.36 &   518 $\pm$ 11 \\ \hline
       86 &      1076.86 &       -84.21 & 20120421 &          239 &    7.48 $\pm$ 1.01 & 0.81 $\pm$ 0.23 &    486 $\pm$ 9 \\ \hline
       87 &     -1107.69 &        27.36 & 20120121 &          239 &  87.26 $\pm$ 14.96 & 1.15 $\pm$ 0.42 &    529 $\pm$ 8 \\ \hline
       88 &     -1137.79 &       -80.01 & 20120121 &          243 &    4.61 $\pm$ 0.69 & 0.53 $\pm$ 0.19 &    433 $\pm$ 7 \\ \hline
       89 &     -1059.30 &        89.12 & 20120423 &          246 &   37.74 $\pm$ 9.51 & 1.12 $\pm$ 0.32 &    492 $\pm$ 9 \\ \hline
       90 &      1133.61 &        29.86 & 20120719 &          249 &   10.73 $\pm$ 2.78 & 1.26 $\pm$ 0.53 &   699 $\pm$ 17 \\ \hline
       91 &      1094.80 &        76.04 & 20120719 &          252 &  -32.13 $\pm$ 6.69 & 1.53 $\pm$ 0.64 &   575 $\pm$ 16 \\ \hline
       92 &      -364.74 &      1059.90 & 20120121 &          255 &    9.64 $\pm$ 1.61 & 0.96 $\pm$ 0.35 &    531 $\pm$ 8 \\ \hline
       93 &       340.99 &      -998.73 & 20130717 &          265 &     9.4 $\pm$ 1.83 & 1.12 $\pm$ 0.49 &   528 $\pm$ 10 \\ \hline
       94 &     -1111.53 &       -73.92 & 20120410 &          272 & 298.17 $\pm$ 44.38 & 1.22 $\pm$ 0.52 &    436 $\pm$ 7 \\ \hline
       95 &      -861.38 &      -750.02 & 20120410 &          275 &   19.62 $\pm$ 2.79 & 0.99 $\pm$ 0.42 &    417 $\pm$ 5 \\ \hline
       96 &       583.79 &      -974.75 & 20120121 &          275 &    7.17 $\pm$ 1.41 & 1.04 $\pm$ 0.39 &   663 $\pm$ 12 \\ \hline
       97 &      1096.67 &       -67.21 & 20120421 &          275 &   11.61 $\pm$ 1.24 & 0.95 $\pm$ 0.27 &    448 $\pm$ 5 \\ \hline
       98 &       597.43 &      -975.40 & 20120121 &          281 &   54.77 $\pm$ 10.0 &  0.89 $\pm$ 0.3 &   653 $\pm$ 11 \\ \hline
       99 &       613.27 &      -918.13 & 20120810 &          281 &   20.03 $\pm$ 3.81 &  0.99 $\pm$ 0.4 &   619 $\pm$ 11 \\ \hline
      100 &     -1095.74 &        66.67 & 20120423 &          297 &    9.52 $\pm$ 1.96 & 0.96 $\pm$ 0.27 &    482 $\pm$ 7 \\ \hline
      101 &       559.78 &     -1031.05 & 20120121 &          297 &    3.67 $\pm$ 0.61 & 0.61 $\pm$ 0.22 &   593 $\pm$ 12 \\ \hline
      102 &      1030.30 &       206.40 & 20120121 &          314 &  -13.52 $\pm$ 1.79 & 1.17 $\pm$ 0.43 &   511 $\pm$ 17 \\ \hline
      103 &      1113.96 &       -63.69 & 20120421 &          317 &    7.43 $\pm$ 0.78 & 0.88 $\pm$ 0.26 &   488 $\pm$ 11 \\ \hline
      104 &       715.05 &      -781.73 & 20130410 &          320 &   15.01 $\pm$ 2.34 &  1.18 $\pm$ 0.5 &   532 $\pm$ 16 \\ \hline
      105 &       984.60 &      -432.80 & 20140102 &          327 &    70.0 $\pm$ 6.63 &  1.0 $\pm$ 0.42 &   338 $\pm$ 33 \\ \hline
      106 &     -1116.05 &        36.56 & 20120120 &          327 &   -16.65 $\pm$ 1.8 & 1.65 $\pm$ 0.69 &   388 $\pm$ 13 \\ \hline
      107 &      1131.76 &       -42.93 & 20120421 &          336 &    8.45 $\pm$ 0.83 & 0.85 $\pm$ 0.25 &   443 $\pm$ 43 \\ \hline
      108 &       906.47 &       552.07 & 20120719 &          340 &   57.12 $\pm$ 6.52 & 1.08 $\pm$ 0.45 &    428 $\pm$ 7 \\ \hline
      109 &      1135.61 &       100.60 & 20120719 &          349 &   21.41 $\pm$ 3.56 & 1.21 $\pm$ 0.51 &   632 $\pm$ 17 \\ \hline
      110 &       992.61 &       476.86 & 20120719 &          362 &   22.41 $\pm$ 2.43 &  0.96 $\pm$ 0.4 &    429 $\pm$ 6 \\ \hline
      111 &       935.90 &       647.84 & 20120810 &          365 & -102.21 $\pm$ 9.29 &  1.2 $\pm$ 0.49 &    388 $\pm$ 4 \\ \hline
      112 &     -1094.45 &       -56.87 & 20120121 &          417 &    9.39 $\pm$ 0.98 & 0.63 $\pm$ 0.24 &    530 $\pm$ 9 \\ \hline
      113 &       429.46 &     -1048.22 & 20130717 &          430 &   16.81 $\pm$ 1.96 & 0.97 $\pm$ 0.41 &   525 $\pm$ 11 \\ \hline
      114 &      -317.00 &      1032.62 & 20120120 &          453 &   19.62 $\pm$ 1.62 &  1.17 $\pm$ 0.5 &   403 $\pm$ 11 \\ \hline
      115 &      -666.95 &       794.06 & 20120514 &          466 &   -21.1 $\pm$ 1.35 & 2.61 $\pm$ 0.87 &   399 $\pm$ 11 \\ \hline
      116 &       912.78 &       532.07 & 20120719 &          475 &  116.17 $\pm$ 8.96 & 1.05 $\pm$ 0.44 &    407 $\pm$ 7 \\ \hline
      117 &      -322.68 &      1050.09 & 20120120 &          475 &  -15.88 $\pm$ 2.23 & 1.78 $\pm$ 0.76 &   738 $\pm$ 17 \\ \hline
      118 &       822.03 &      -708.55 & 20120423 &          479 &   11.75 $\pm$ 1.81 & 0.74 $\pm$ 0.22 &   567 $\pm$ 10 \\ \hline
      119 &       596.26 &      -935.79 & 20120120 &          524 &  -27.88 $\pm$ 2.64 &  1.67 $\pm$ 0.7 &   549 $\pm$ 15 \\ \hline
      120 &     -1077.38 &      -109.12 & 20120121 &          585 &    81.73 $\pm$ 5.7 &  1.36 $\pm$ 0.5 &    517 $\pm$ 9 \\ \hline
\enddata
\end{deluxetable*}

\pagebreak
\bibliographystyle{aasjournal}
\bibliography{stat_paper.bib}



\end{document}